\documentclass[aps,prb,reprint,amsmath,floatfix]{revtex4-2}

\usepackage{graphicx}
\usepackage[colorlinks=true,linkcolor=black,citecolor=blue,urlcolor=blue]{hyperref}

\newcommand{\vc}[1]{\boldsymbol{#1}}

\begin{document}

\title{Emergent transverse-field Ising model in 
$\boldsymbol{d}^\mathbf{4}$ spin-orbit Mott insulators}

\author{Ji\v{r}\'{\i} Chaloupka}

\affiliation{Department of Condensed Matter Physics, Faculty of Science,
Masaryk University, Kotl\'a\v{r}sk\'a 2, 61137 Brno, Czech Republic}

\begin{abstract}
Mott-insulating transition metal oxides containing $t_{2g}^4$ ions with strong
spin-orbit coupling were recently demonstrated to display unusual magnetism
due to a dynamical mixing of the low-energy multiplet states via exchange
processes. Here we derive exchange interactions in the situation where a
tetragonal or trigonal crystal field selects a single relevant excited state
on top of the singlet ionic ground state, producing an effective
spin-$\frac12$. We show that these moments universally obey antiferromagnetic
transverse-field Ising model (TFIM) with an intrinsic transverse field
generated by the splitting of the two ionic singlets. Using Ru$^{4+}$ as a
example ion, we provide quantitative estimates of the exchange and illustrate
the emergent TFIM physics based on phase diagrams and excitation spectra
obtained for several 2D lattices -- square, honeycomb, as well as for the
frustrated triangular lattice.
\end{abstract}

\date{\today}

\maketitle



\textit{Introduction.}
Transverse-field Ising model introduced by de Gennes in 1960s \cite{deG63}
ranks among the most prominent example systems to study quantum criticality
\cite{Sac11,Dut15}. Its Ising part belongs to a few lattice models in
statistical physics for which exact solutions are available
\cite{Isi25,Ons44,Kau49a,Kau49b,Wan50,Wan73} and gained popularity as a
prototype model to capture collective behavior not only of localized spins in
magnets but also in a much broader context \cite{Cas09,Sor14}. Ising model
itself represents a classical problem. By adding transverse magnetic field,
the quantum nature of spins comes into play, leading to a quantum critical
behavior reflecting a competition of Ising interactions and Zeeman energy. The
model has been studied in various settings, a particularly appealing case is
the antiferromagnetic (AF) TFIM on a triangular lattice combining quantum
criticality with frustration and exhibiting Berezinskii-Kosterlitz-Thouless
(BKT) transitions \cite{Moe00,Moe01,Isa03}. 

Realization of TFIM in magnetic materials is limited by the requirement of
strong uniaxial anisotropy of exchange interactions and their suitable
strength with respect to accessible magnetic fields. Simple anisotropic
ferromagnets were considered since the early days \cite{Sti73}, however a
definitive experimental demonstration came only in 2010 with
\mbox{CoNb$_2$O$_6$} acting as 1D TFIM chain in neutron scattering
\cite{Col10}. 
The case of frustrated AF TFIM is yet more delicate. A~promising route was
recently suggested by $4f$ triangular-lattice compound \mbox{TmMgGaO$_4$}
where signatures of BKT physics were observed \cite{Li20,Hu20,Dun21}. Here
the lowest two levels of Tm$^{3+}$ ions form a well-isolated pair of singlets
hosting effective spin-$\frac 12$. Thanks to large spin-orbit coupling (SOC)
and hence large orbital component of these moments, they are subject to
strongly anisotropic interactions. The second key ingredient is the
crystal-field (CF) induced splitting of the two singlets that plays a~role of
an \textit{intrinsic} transverse field \cite{She19,Liu19,Che19,Liu20}.
In this Letter we show that $4d^4$ and $5d^4$ Mott insulators with large SOC
\cite{Tak21} may be even more convenient platform for TFIM utilizing a similar
mechanism. As we demonstrate by explicitly deriving the exchange model,
at sufficiently large negative CF splitting of $t_{2g}$ levels, the
interactions are of AF Ising-type enforced by the very structure of the $d^4$
ionic states. This promises a realization of TFIM on various 2D lattices at
larger energy scales and corresponding characteristic temperatures as compared
to $4f$ systems (exchange strength reaching tens of meV compared to
$\approx 1\:\mathrm{meV}$ for \mbox{TmMgGaO$_4$} \cite{Li20}), with the
intrinsic transverse field potentially tunable by straining.



\begin{figure}[t!b]
\includegraphics[scale=1.0]{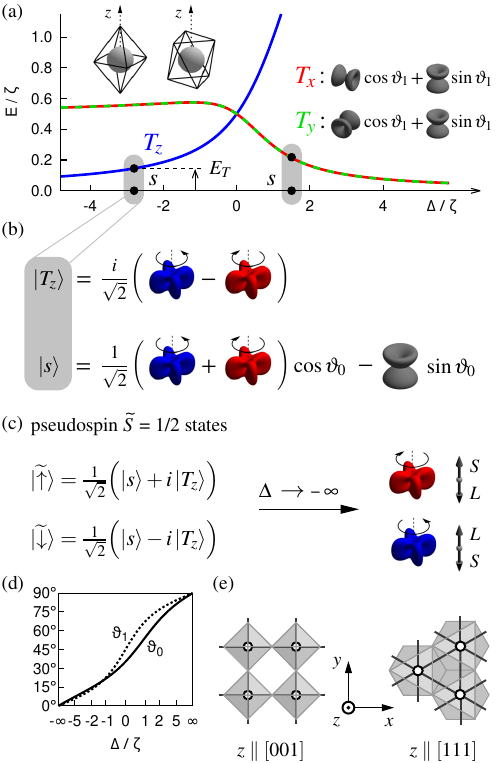}
\caption{(a)~Low-energy levels of $t_{2g}^4$ ion depending on tetragonal or
trigonal CF splitting $\Delta$. The energies are plotted relative to the
ionic ground-state level. Within $LS$ coupling scheme, total $L=1$ and $S=1$
of the two holes added to the $t_{2g}^6$ configuration are combined by SOC
into ground-state singlet $s$, a triplet of states $T_{x,y,z}$ degenerate at
$\Delta=0$, and a quintuplet at higher energies that may be ignored. In
point-charge model, negative $\Delta$ corresponds to the indicated 
tetragonal/trigonal elongation of metal-O$_6$ octahedra
(in reality it is influenced also by the Madelung potential contributed by
more distant ions and covalency effects which may break this simple
correspondence).
(b)~At large enough $\Delta<0$, the pair $s$ and $T_z$ forms the basis of the
low-energy model. The corresponding wave functions (for the tetragonal case) are
represented by shapes of the respective hole densities resolved according to
$S_z$ and $L_z$. Spin is indicated by color (red: up along $z$, blue: down),
orbital angular momentum by an arrow. The remaining $T_{x,y}$ are shown in the
inset of (a).
(c)~Linear combinations of $s$ and $T_z$ forming the basis of the pseudospin
$\widetilde{S}=\frac12$ model. For large negative $\Delta$ they converge to
fully polarized states with antiparallel $L$ and $S$.
(d)~$\Delta$-dependent angles $\vartheta_{0,1}$ entering the wave functions.
(e)~$xyz$ reference frame with $z$ being the out-of-plane axis. 
}\label{fig:intro}
\end{figure}

\textit{TFIM imposed by spin-orbital structure of ionic states.}
We first briefly review the multiplet structure of relevant ions such as
\mbox{Ru$^{4+}$} or \mbox{Ir$^{5+}$} with $d^4$ valence configuration limited
to $t_{2g}$ orbitals. Large SOC in these ions forms non-magnetic $J=0$ singlet
ground state $s$ and low-energy $J=1$ triplet excitations $T_x$, $T_y$, $T_z$
separated from the ground-state level by energy $\zeta/2$ with $\zeta$
denoting the single-electron SOC strength \cite{Tak21}. These states may serve
as a basis for an effective model exhibiting quantum critical behavior due to
the competition of exchange and triplet energy cost \cite{Kha13,Tak21}. The
essential control parameter here is the tetragonal or trigonal CF splitting
$\Delta$ of $t_{2g}$ orbital levels relevant for 2D lattices of corner-sharing
or edge-sharing \mbox{metal-O$_6$} octahedra, respectively. Nonzero $\Delta$
splits the $J=1$ triplet and strongly modifies the ionic excitation energies
as plotted in Fig.~\ref{fig:intro}(a) and thoroughly discussed in
\cite{noteSM}.
Earlier works analyzed the situation with no triplet splitting 
\cite{Kha13,Ani19,Cha19} and partially the positive-$\Delta$
case \cite{Jai17,Sou17}.
Here we focus on the so far unexplored negative-$\Delta$ case with
$|\Delta|\gtrsim\zeta$, the relevant set of basis states thus gets reduced to
a non-Kramers doublet consisting of the ionic ground state $s$ and one of the
triplets $T_z$ selected by the out-of-plane CF axis direction $z$ [see
Fig.~\ref{fig:intro}(a),(e)]. The internal structure of these $t_{2g}^4$
states sketched in Fig.~\ref{fig:intro}(b) is best appreciated when expressing
them via eigenstates $|L_z,S_z\rangle$ of the total effective $t_{2g}$ orbital
momentum $L=1$ and total spin $S=1$ carried by the two $t_{2g}$ holes:
\begin{align}
|s\rangle   &= \tfrac1{\sqrt2}(|\!+\!1,-1\rangle+|\!-\!1,+1\rangle) \cos\vartheta_0 
              -|0,0\rangle \sin\vartheta_0 \,, \notag \\
|T_z\rangle &= \tfrac{i}{\sqrt2}(|\!+\!1,-1\rangle-|\!-\!1,+1\rangle) \,.
\end{align}
The auxiliary angle $\vartheta_0$ is given by 
$\tan 2\vartheta_0 = 2\sqrt2/(1-2\delta)$ with $\delta=\Delta/\zeta$
and vanishes in $\Delta\rightarrow -\infty$ limit [see Fig.~\ref{fig:intro}(d)].
The splitting of $s$ and $T_z$ levels equals
$E_T=\frac14\zeta[\sqrt{(1-2\delta)^2+8}-(1-2\delta)]$, vanishing as 
$\zeta/(1-2\delta)$.

The exchange interactions between $d^4$ ions in the above regime can be
obtained by standard second-order perturbation theory in electron hopping
resulting in a model for hardcore bosons $s$ and $T_z$. These are subject to
the local constraint $n_s+n_{T_z}=1$, where $n_s=s^\dagger s$ and
$n_{T_z}=T_z^\dagger T_z^{\phantom{\dagger}}$ count bosons on a given site. 
The model becomes particularly transparent if formulated in
terms of a pseudospin $\widetilde{S}=\frac12$ based on the linear
combinations $|\widetilde{\uparrow}\rangle, |\widetilde{\downarrow}\rangle =
\tfrac1{\sqrt 2}(|s\rangle \pm i|T_z\rangle)$:
\begin{align}
|\widetilde{\uparrow}\rangle &= 
c^2 |\!-\!1,+1\rangle -s^2 |\!+\!1,-1\rangle - \sqrt2\, cs\, |0,0\rangle \,, \notag \\
|\widetilde{\downarrow}\rangle &= 
c^2 |\!+\!1,-1\rangle -s^2 |\!-\!1,+1\rangle - \sqrt2\, cs\, |0,0\rangle 
\end{align}
with $c=\cos\tfrac{\vartheta_0}2$ and $s=\sin\tfrac{\vartheta_0}2$. This
choice is motivated by the $\Delta\rightarrow -\infty$ limit depicted in
Fig.~\ref{fig:intro}(c) where $|\widetilde{\uparrow}\rangle$,
$|\widetilde{\downarrow}\rangle$ correspond to fully polarized states
$|\!-\!1,+1\rangle$, $|\!+\!1,-1\rangle$ with strictly antiparallel
out-of-plane $L_z=-S_z$. Moreover, the pseudospin carries Van Vleck-type
magnetic moment which is purely out-of-plane with large $g_z=6\cos\vartheta_0$
and zero $g_{xy}$ \cite{noteSM}.
On the operator level, this change of the basis is expressed via the
correspondence relations
$\widetilde{S}_x = \tfrac12 -n_{T_z}$,
$\widetilde{S}_y = \tfrac12(s^\dagger T_z + T_z^\dagger s)$,
$\widetilde{S}_z = -\tfrac{i}2(s^\dagger T_z - T_z^\dagger s)$.
As a consequence, the level splitting $E_T$ translates to a transverse field
$h=E_T$ in the pseudospin formulation. The form of the exchange interactions
can be easily anticipated by considering the $\Delta\rightarrow -\infty$ limit
in Fig.~\ref{fig:intro}(c).
Since the two virtual electronic hoppings generating second-order exchange can
only change the ionic spin component $S_z$ by $\Delta S_z=0,\pm 1$, the states
$|\widetilde{\uparrow}\rangle$, $|\widetilde{\downarrow}\rangle$ with 
$S_z=\pm 1$ cannot be connected and the exchange is strictly of Ising
$\widetilde{S}_z\widetilde{S}_z$ type in this limit. A full derivation for
general $\Delta\lesssim -\zeta$ gives the pseudospin model
\begin{equation}\label{eq:HS}
\mathcal{H}_{\widetilde{S}}=
\sum_{\langle ij\rangle} \left(
J_z \widetilde{S}^z_i \widetilde{S}^z_j 
+ J_x \widetilde{S}^x_i \widetilde{S}^x_j 
+ J_y \widetilde{S}^y_i \widetilde{S}^y_j \right)
-(h+\delta h) \sum_i \widetilde{S}^x_i
\end{equation}
with dominant $J_z$ and $h=E_T$, supplemented by minor $J_x$, $J_y$, 
$\delta h$. In contrast to $J_z$, the latter exchange parameters 
only arise due to the small common parts of $|\widetilde{\uparrow}\rangle$, 
$|\widetilde{\downarrow}\rangle$ and as such they are
proportional to $\sin^2\vartheta_0$. As demonstrated later,
they quickly drop when entering the $\Delta\lesssim-\zeta$ regime.
Detailed exchange expressions for both $180^\circ$ bonds (corner-sharing
\mbox{metal-O$_6$} octahedra) and $90^\circ$ bonds (edge-sharing) as well as
the connection to the hardcore boson formulations are given in \cite{noteSM}.
Note that due to the omission of the bond-directional states $T_{x,y}$,
the interactions are identical for all bond directions.
Neglecting the minor contributions in Eq.~\eqref{eq:HS}, we arrive at the
final minimal model which takes the form of transverse-field Ising model
\begin{equation}\label{eq:HTFIM}
\mathcal{H}_\mathrm{TFIM}=
J_z \sum_{\langle ij\rangle}
\widetilde{S}^z_i \widetilde{S}^z_j 
-h \sum_i \widetilde{S}^x_i \,.
\end{equation}
Let us emphasize that this minimal model is imposed solely by the internal
structure of the ionic states at sufficiently negative $\Delta/\zeta$, hence
the mechanism is universal for any lattice. 



\begin{figure}[tb]
\includegraphics[scale=1.0]{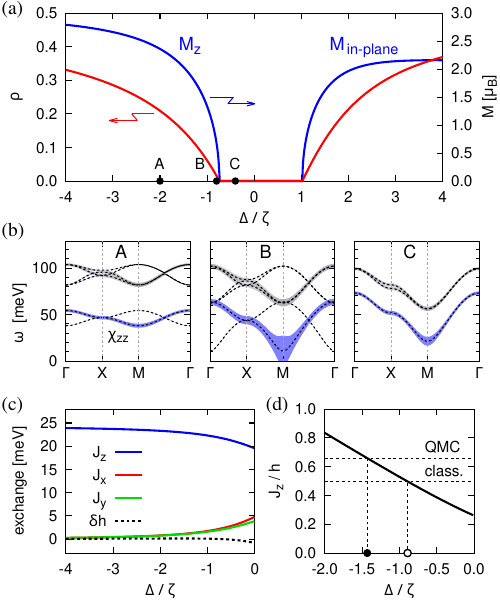}
\caption{(a)~Variational phase diagram of the full $s$-$\vc T$ model for the
square lattice obtained using $U=3\:\mathrm{eV}$, $J_\mathrm{H}=0.5\:\mathrm{eV}$,
$\zeta=0.15\:\mathrm{eV}$ (roughly corresponding to \mbox{Ru$^{4+}$}), 
and hopping $t=0.14\:\mathrm{eV}$. At large enough
negative/positive $\Delta$, condensate of $\vc T$-bosons with the density
$\rho$ is present and carries out-of-plane/in-plane AF moment. 
(b)~Dynamic magnetic susceptibility calculated by LFWT at selected points in 
(a) and separated into $zz$ component (blue) and in-plane $xx+yy$ part (gray).
Line thickness scales with the intensity, dashed lines indicate the
dispersions of excitations. Triangular Brillouin-zone path including
high-symmetry points $\Gamma=(0,0)$, $X=(\pi,0)$, and $M=(\pi,\pi)$ is used.
Identical $zz$ component spectra are obtained also by LFWT limited to $s$ and
$T_z$ bosons only.
(c)~Plot of the interaction parameters of the effective pseudospin-$\frac12$ model
\eqref{eq:HS} showing the clear dominance of $J_z$.
(d)~Parameter ratio $J_z/h$ of the effective model plotted as function of
$\Delta/\zeta$. The QCP position on $\Delta/\zeta$ axis is estimated by
mapping the critical ratio $(J_z/h)_\mathrm{crit}$ of TFIM determined either
classically ($0.5$) or using precise QMC result ($\approx 0.657$) back to
$\Delta/\zeta$.
}\label{fig:sqr}
\end{figure}

\textit{Exchange parameters, phase diagrams, and excitations.}
In this section, we illustrate the emergence of TFIM as a low-energy magnetic
model by exploring phases and excitations obtained using an exchange model
including all four low-energy states $s$, $T_{x,y,z}$ in the local basis. This
model, to be called (full) \mbox{$s$-$\vc T$} model in the following, has the
advantage to be applicable at any $\Delta/\zeta$ and allows us to study the
crossover to the $\Delta\lesssim -\zeta$ regime of interest. As we show below,
the main features in this regime can indeed be understood and reproduced by
simple AF TFIM. Its connection to the full \mbox{$s$-$\vc T$} model is
provided by projection onto $s$ and $T_z$, which transforms the \mbox{$s$-$\vc
T$} model into the pseudospin-$\frac12$ model of Eq.~\eqref{eq:HS}. On the
way, we will also give quantitative hints on the TFIM parameters targeting
Ru$^{4+}$ compounds by the particular choice of Hubbard repulsion $U$, Hund's
coupling $J_\mathrm{H}$, and SOC strength $\zeta$.
The \mbox{$s$-$\vc T$} model Hamiltonian was obtained by second-order
perturbation theory and encompasses a large number of bond terms involving
hardcore bosons $s$, $T_{x,y,z}$, following the general structure presented in
\cite{noteSM}. Due to this complexity, it cannot be given explicitly here (see
\cite{Kha13} for a simpler version with $\Delta/\zeta=J_\mathrm{H}/U=0$), but
we use it in full to determine the variational phase diagram using the trial
product state 
\begin{equation}\label{eq:Psitrial}
|\Psi_\mathrm{trial}\rangle = \!\! \prod_{i \in \text{sites}} 
\Bigl(\sqrt{1-\rho_i}\; s^\dagger +\sqrt{\rho_i} \!\!
{\sum_{\alpha=x,y,z}} 
d^*_{i\alpha} T_{i\alpha}^\dagger \Bigr) |\mathrm{vac}\rangle
\end{equation}
and to calculate corresponding harmonic excitations using linear flavor
wave theory (LFWT) \cite{noteSM,Pap84,Pap88,Chu90,Jos99,Som01}. 
$|\Psi_\mathrm{trial}\rangle$ of Eq.~\eqref{eq:Psitrial} enables to capture various forms
of magnetically ordered states linked to a condensation of hardcore vector
bosons $\vc T$ as well as the paramagnetic state where $\vc T$ remain
uncondensed. In the former case, the site-dependent variational parameters
$\rho_i$ (scalars) and $\vc d_i$ (unit vectors) determine the condensate
density and magnetic structure, respectively. The associated excitation
spectrum contains magnon-like modes (fluctuations in $\vc d$) and amplitude
mode (oscillations of the condensate density $\rho$). In the latter
paramagnetic case, we find a trivial minimum with all $\rho_i=0$ and 
excitations being carried directly by bosons $\vc T$.

As a first example, Fig.~\ref{fig:sqr} gives an overview for a square lattice
with straight $180^\circ$ bonds, where the nearest-neighbor hopping $t$
connects diagonally a pair of $t_{2g}$ orbitals active on a given bond
\cite{Tak21}. The phase diagram shown in Fig.~\ref{fig:sqr}(a) contains a
window of paramagnetic (PM) phase around $\Delta/\zeta=0$, separated by
quantum critical points (QCP) at $\Delta\approx \pm\zeta$ from two condensed
phases. Both are characterized by AF ordered Van Vleck moments but their
nature strongly differs. The positive-$\Delta$ case with in-plane moments can
be described by a pseudospin-1 model with predominantly XY-type of
interactions and has been discussed in the context of \mbox{Ca$_2$RuO$_4$}
\cite{Jai17,Sou17,noteSM} which was estimated to have 
$\Delta/\zeta\approx 1.5$ \cite{Jai17}.
In contrast, our negative-$\Delta$ case of interest is captured by the above
pseudospin-$\frac12$ TFIM. In this language, the pseudospins in the PM phase
are fully aligned by the in-plane transverse field $h=E_T$, while beyond
QCP they develop staggered out-of-plane component supported by $J_z>0$. Due to
zero in-plane $g$-factor, only the AF out-of-plane component carries magnetic
moment.

\begin{figure}[tb]
\includegraphics[scale=1.0]{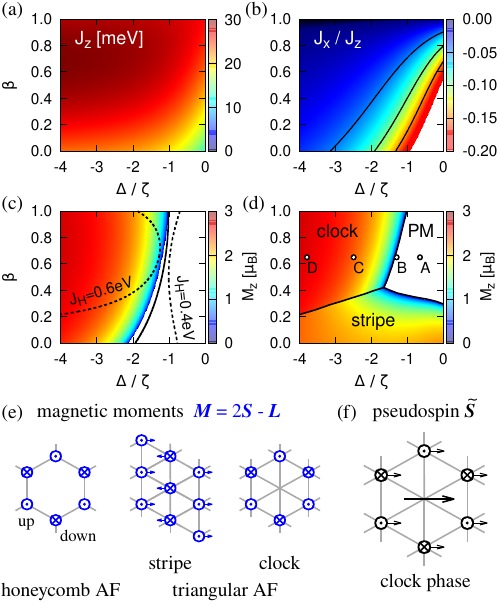}
\caption{(a)~Effective exchange parameter $J_z$ for the 90$^\circ$ bonding
geometry and $U=3\:\mathrm{eV}$, $J_\mathrm{H}=0.5\:\mathrm{eV}$,
$\zeta=0.15\:\mathrm{eV}$. The parameter $\beta$ interpolates linearly between
metal-O-metal hopping $t$ and direct hopping $t'$, namely $t=(1-\beta)\times
0.3\:\mathrm{eV}$, $t'=\beta\times 0.3\:\mathrm{eV}$.
(b)~Relative strength of $J_x$ ($\approx J_y$) compared to the dominant $J_z$.
(c)~Variational phase diagram of the full $s$-$\vc T$ model on the honeycomb
lattice. Major part is taken by AF phase with out-of-plane moments, the white
area corresponds to the PM phase. The (classical) phase boundary of the
resulting TFIM (i.e. pseudospin-$\frac12$ model with $J_{x,y}$ neglected) is
shown by the solid line. Dashed lines illustrate its trend for varying
$J_\mathrm{H}$.
(d)~Variational phase diagram for the triangular lattice. Using the
critical ratios for TFIM from QMC \cite{Wan17} together with our TFIM
parameters, the estimated phase boundary gets shifted horizontally by about
$0.3$ to the left in (c) and by $1.2$ in (d).
(e)~Magnetic patterns of the phases in (c),(d).
(f)~Clock-phase pattern in the pseudospin-$\frac12$ representation. The
central site carries a saturated in-plane moment which is completely 
hidden in the magnetic pattern due to the zero in-plane $g$-factor.
}\label{fig:tri}
\end{figure}

The TFIM picture is confirmed by the excitation spectra in
Fig.~\ref{fig:sqr}(b). Near the QCP, the dispersion of the low-energy
excitations probed by $\chi_{zz}$ susceptibility softens at the AF
momentum $M=(\pi,\pi)$, deeper in the AF phase they become flat and the gap
saturates, which is consistent with the expected Ising-type excitation at
constant $\omega=2J_z$ contrasting to the magnon-like excitations for positive
$\Delta$ \cite{noteSM}. The proximity to TFIM is illustrated by the evaluated
parameters of the \mbox{pseudospin-$\frac12$} model \eqref{eq:HS} presented in
Fig.~\ref{fig:sqr}(c). For $\Delta\lesssim -\zeta$, the dominant AF $J_z$
quickly saturates at the infinite-$\Delta$ value $J_z\approx(5-7\eta)\,t^2/U$
with $\eta=J_H/U$ \cite{noteSM}, and is accompanied by tiny in-plane
$J_{x,y}$. Finally, Fig.~\ref{fig:sqr}(d) combines $J_z$ of
Fig.~\ref{fig:sqr}(c) and $h=E_T$ found in Fig.~\ref{fig:intro}(a) into the
ratio $J_z/h$ that is the decisive parameter of TFIM and can be used to
estimate the critical value of $\Delta/\zeta$. The value
$(\Delta/\zeta)_\mathrm{crit}\approx -0.9$ based on
$(J_z/h)_\mathrm{crit}=0.5$ obtained by treating TFIM classically roughly
agrees with our variational result for the full $s$-$\vc T$ model in
Fig.~\ref{fig:sqr}(a). It gets corrected towards more negative
$(\Delta/\zeta)_\mathrm{crit}\approx -1.4$ when using the precise value
$(J_z/h)_\mathrm{crit}\approx 0.657$ obtained by QMC \cite{Hua20}.

Similar analysis is performed for $90^\circ$ bond geometry that occurs in e.g.
honeycomb or triangular lattices with edge-sharing octahedra. Here two major
hopping channels active on \mbox{metal$_2$-O$_2$} plaquettes have to be
simultaneously considered -- bonding paths via oxygen ions and a direct
overlap of $d$ orbitals \cite{Tak21}. The former hopping with amplitude $t$
connects off-diagonally a~bond-dependent pair of $t_{2g}$ orbitals while the
complementary $t_{2g}$ orbital is subject to direct hopping $t'$. Despite the
completely different hopping rules as compared to the $180^\circ$ case, the
pseudospin-$\frac12$ interactions plotted in Figs.~\ref{fig:tri}(a),(b) again
feature dominant AF Ising $J_z$ accompanied by minor in-plane $J_{x,y}$, in
accord with the general conclusions of the previous section. For large
negative $\Delta$, the value of $J_z$ approaches 
$J_z\approx \frac49\, [(7-20\eta)\,t^2+(2+8\eta)\,tt'+(4-8\eta)\,t'^2]/U$.

Variational phase diagram for the non-frustrated honeycomb lattice presented
in Fig.~\ref{fig:tri}(c) and the corresponding excitation spectrum
\cite{noteSM} show similar behavior as for the square lattice. Much richer is
the case of the frustrated triangular lattice. The phase diagram shown in
Fig.~\ref{fig:tri}(d) contains two condensed phases with non-trivial patterns
depicted in Fig.~\ref{fig:tri}(e),(f). One of them appears in so-called clock
phase familiar from the studies of TFIM on triangular lattice
\cite{Moe01,Isa03}. Here the frustration of pseudospins is resolved by a
formation of honeycomb AF pattern of the out-of-plane components to satisfy
the Ising interactions and a simultaneous alignment of the in-plane components
with the transverse field. At the remaining sites, the pseudospins are
strictly in-plane, avoiding the Ising interactions and conforming fully to the
transverse field. The other pattern -- of stripy type -- is specific to our
$s$-$\vc{T}$ model and is not captured by the pseudospin-$\frac12$ TFIM
because of the participation of $T_{x,y}$ in the condensate. Based on the
extended nature of $4d$ and $5d$ orbitals, the regime $t'\gtrsim t$ can be
expected, making the clock phase more relevant. 

\begin{figure*}[tb]
\includegraphics[scale=1.0]{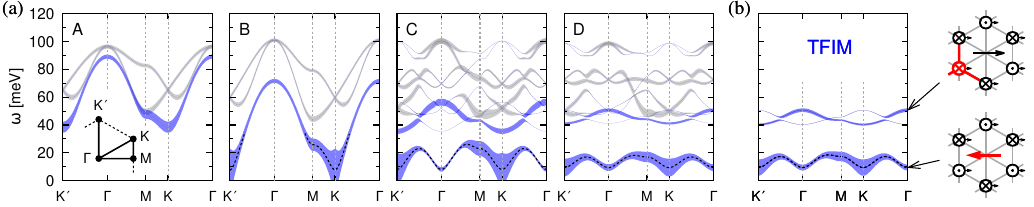}
\caption{(a)~Magnetic excitation spectra for the selected points indicated in
Fig.~\ref{fig:tri}(d) as obtained by LFWT applied to $s$-$\vc T$ model. The
intensity is represented by line thickness as in Fig.~\ref{fig:sqr}(b) with
blue indicating the out-of-plane $zz$ component and gray the in-plane
components. The inset shows the Brillouin zone path.
(b)~$zz$ susceptibility for the point D calculated using linear spin-wave
theory applied to the corresponding TFIM with $J_z\approx 30\:\mathrm{meV}$
and $h\approx 19\:\mathrm{meV}$. Cartoons capture the two distinct excitations
-- low-energy fluctuations of the in-plane pseudospins and high-energy
Ising-type excitations.
}\label{fig:triexc}
\end{figure*}

Interestingly, the clock pattern of pseudospins on triangular lattice is
obscured by zero in-plane $g$-factor, giving rise to a static magnetic
pattern identical to that of the honeycomb AF phase. However, the excitation
spectra reveal a fundamental difference to the latter case.
Figure~\ref{fig:triexc}(a) shows the evolution of the magnetic excitations
when crossing the PM/clock phase boundary. These are again obtained using the
full $s$-$\vc T$ model, but -- as demonstrated by Fig.~\ref{fig:triexc}(b) 
-- the relevant $\chi_{zz}$ is perfectly reproduced also by the pseudospin
dynamics within the corresponding TFIM. Approaching the PM/clock 
boundary, the excitations soften at the characteristic momentum
$K=2\pi(\frac1{\sqrt3},\frac13)$ of the honeycomb AF pattern formed after
entering the clock phase. Here a separation of energy scales occurs. The
high-energy part of the spectrum is represented by a flat Ising-type
excitation at $\omega\approx 2J_z$ encountered previously and linked here to a
pseudospin flip taking place in the honeycomb AF structure. This excitation
is complemented by a dispersing low-energy mode (energy scale $h$) that
involves rotations of in-plane pseudospins.
Its dispersion, soft near $K$, is approximately given by
$\omega_{\vc q} \approx h[1-(h/h_c)|\gamma_{\vc q}|^2 ]$ with
$\gamma_{\vc q}=2\cos(\frac{\sqrt{3}}2 q_x)+\exp(i \frac32 q_y)$
and $h_c=\frac32 J_z$ denoting the critical transverse field.
 Remarkably, this intense magnetic
excitation stems from the moments that are magnetically invisible in the
static pattern. 
In contrast to the triangular lattice case, the honeycomb spectrum
\cite{noteSM} hosts only the Ising-type excitation and its low-energy part is
empty.



\textit{Conclusions.}
We presented a detailed theoretical account on the exchange interactions in
$4d^4$ and $5d^4$ spin-orbit Mott insulators in the regime of negative
tetragonal or trigonal crystal field $\Delta$. As illustrated by the
corresponding phase diagrams and excitations for several 2D lattices, the
low-energy magnetism can be well captured by AF transverse-field Ising model
involving effective \mbox{spins-$\frac12$}. Being based on $d$ valence
electrons, the emergent TFIM features convenient energy scales in the range of
tens of meV. The transverse field is intrinsic, generated by CF itself, and is
therefore sensitive to strain control. Robust Ising-type interactions are
imposed by the internal spin-orbital structure of the $d^4$ ionic states, and
as such they are generic to both $180^\circ$ and $90^\circ$ bonding
geometries. This universality of TFIM description is in strong contrast to the
much different behavior of singlet-triplet models obtained for
$\Delta/\zeta=0$ in these two bonding-geometry cases \cite{Kha13,Ani19,Cha19}.
The radical change of the magnetic model when varying $\Delta/\zeta$ is an
excellent illustration of the richness of the exchange interactions among $4d$
and $5d$ ions brought about by the complex structure of the low-energy ionic
states. Apart from promising an identification/engineering of TFIM in the
family of $4d$ and $5d$ correlated oxides, the proposed scenario also
motivates the study of related theoretical issues. For example, the
calculations suggest the dominant Ising exchange to be accompanied by small
interactions between transverse components of effective spins. Their influence
on the BKT behavior of TFIM is an interesting open problem.


\textit{Acknowledgments.}
Stimulating discussions with Giniyat Khaliullin are gratefully acknowledged.
This work was supported by Czech Science Foundation (GA\v{C}R) under
Project No.~GA22-28797S.





\onecolumngrid
\renewcommand\thesection{\Alph{section}}
\renewcommand\thefigure{S\arabic{figure}}

\renewcommand{\topfraction}{1.0}
\renewcommand{\textfraction}{0.0}
\renewcommand{\textfloatsep}{10pt}

\clearpage
\setcounter{page}{0}
\setcounter{equation}{0}
\setcounter{section}{0}
\setcounter{figure}{0}

\begin{center}
{\large Supplemental Material for}
\vskip 3mm
{\large\bf Emergent transverse-field Ising model in 
$\boldsymbol{d}^\mathbf{4}$ spin-orbit Mott insulators}
\vskip 5mm
Ji\v{r}\'{\i} Chaloupka
\vskip 1mm
{\it Department of Condensed Matter Physics, Faculty of Science, \\
Masaryk University, Kotl\'a\v{r}sk\'a 2, 61137 Brno, Czech Republic}
\end{center}


\section{Low-energy multiplet states of $t_{2g}^4$ configuration}

To construct the multiplet states forming the low-energy basis, we consider
$t^4_{2g}$ states corresponding to two holes made in the fully populated
$t_{2g}^6$ configuration. Working within the $LS$ coupling scheme, we limit
ourselves to the sector with maximum total spin $S=1$ of the hole pair which
is favored by Hund's coupling. The total $t_{2g}$ orbital momentum of the hole
pair is restricted to $L=1$ in that case. The nine-fold degeneracy of the
$S\!=\!L\!=\!1$ sector is lifted by a~simultaneous action of spin-orbit
coupling (SOC)
\begin{equation}
\mathcal{H}_\mathrm{SOC} = \frac{\zeta}{2S}\,\vc S\cdot\vc L 
\end{equation} 
and tetragonal or trigonal crystal-field (CF) that splits the single-electron
$t_{2g}$ levels. In the tetragonal case with the CF axis being one of the main
octahedral axes, the splitting takes a simple form diagonal in the
conventional $t_{2g}$ orbital basis, for example
$\mathcal{H}_\mathrm{CF} = \tfrac13 {\Delta}
\left( d_{yz}^\dagger d_{yz}^{\phantom{\dagger}}
+ d_{zx}^\dagger d_{zx}^{\phantom{\dagger}}
-2 d_{xy}^\dagger d_{xy}^{\phantom{\dagger}} \right)$ 
if the CF axis coincides with $z$. $\mathcal{H}_\mathrm{CF}$ for the hole
pair may be compactly expressed using the component of total $\vc L$ parallel
to the CF axis:
\begin{equation}\label{eq:HCF}
\mathcal{H}_\mathrm{CF}
=\Delta \left(L_z^2-\tfrac23\right).
\end{equation}
The case of the trigonal splitting is a bit more complicated, involving
so-called $a_{1g}$ a $e_g'$ orbital combinations, but Eq.~\eqref{eq:HCF}
remains valid if we again associate the axis $z$ with the CF axis, i.e.
$z\parallel [111]$ in the octahedral reference frame as shown in Fig.~1 of the
main text.

\begin{figure*}[t]
\includegraphics[scale=1.08]{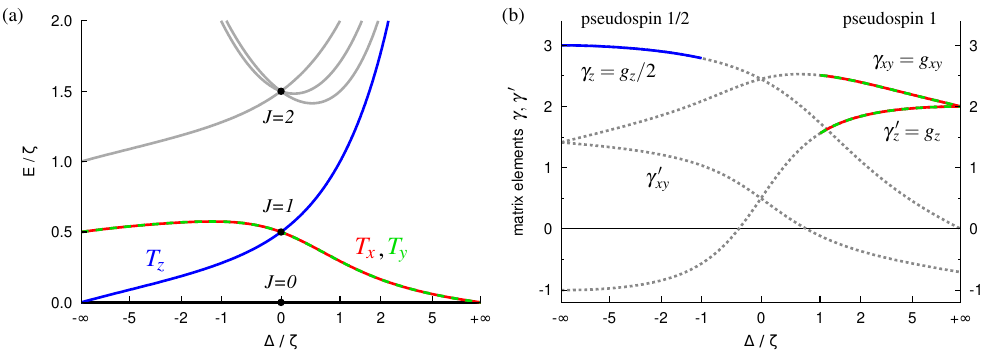}
\caption{(a)~Ionic excitation energies obtained within $LS$ coupling scheme for the
$S=1$, $L=1$ sector of $t_{2g}^4$ favored by Hund's coupling. Plotted are the
energies relative to the ground-state singlet level. At $\Delta/\zeta=0$, the
multiplet states are eigenstates of the total angular momentum
$\vc J=\vc L+\vc S$.
(b)~Factors entering the operators of the magnetic moment in \eqref{eq:SMmx},
\eqref{eq:SMmy}, \eqref{eq:SMmz}. In the $\Delta\lesssim -\zeta$ regime, the
low-energy pair of states $s$ and $T_z$ maps to pseudospin-$\frac12$ with zero
in-plane $g$-factor $g_{xy}=0$ and large out-of-plane $g_z=2\gamma_z$. In the 
$\Delta\gtrsim +\zeta$ regime, the low-energy states $s$, $T_x$, $T_y$ form
pseudospin-1 with the $g$-factors $g_{xy}=\gamma_{xy}$ and $g_z=\gamma'_z$,
both approaching $2$ in the large-$\Delta$ limit where $L$ is quenched leaving only
the spin component active.
}\label{fig:SMlevels}
\end{figure*}

The Hamiltonian $\mathcal{H}_\mathrm{SOC}+\mathcal{H}_\mathrm{CF}$ is easily
diagonalized giving rise to the excitation energies plotted in
Fig.~\ref{fig:SMlevels}(a). In the case of $\Delta/\zeta=0$ where only
$\mathcal{H}_\mathrm{SOC}$ contributes, the energy eigenstates correspond to
$J=0$, $1$, $2$ eigenstates of the total angular momentum $\vc J=\vc L+\vc S$
as evident from the rearrangement 
$\vc S\cdot\vc L=\frac12(\vc J^2-\vc L^2-\vc S^2)$.
In the general case, the energy eigenstates can be conveniently formulated in
terms of the states $|L_z,S_z\rangle$ that simultaneously diagonalize the
out-of-plane components of the total $t_{2g}$ orbital momentum and the total
spin. Thanks to the identical form \eqref{eq:HCF} for both tetragonal and
trigonal splitting, all the following expressions are universal, differing
just in the interpretation of the $z$ axis. 
The singlet ionic ground state $s$ is given by
\begin{equation}
|s\rangle = \tfrac1{\sqrt2}(|\!+\!1,-1\rangle+|\!-\!1,+1\rangle) \cos\vartheta_0 
            -|0,0\rangle \sin\vartheta_0
\end{equation}
with the auxiliary angle obeying $\tan 2\vartheta_0 = 2\sqrt2/(1-2\delta)$
where $\delta=\Delta/\zeta$ measures the relative strength of CF versus SOC.
This angle is plotted in Fig.~1(d) of the main text using the same horizontal
axis as that used in Fig.~\ref{fig:SMlevels}. 
First three excited states forming degenerate $J=1$ triplet at
$\Delta/\zeta=0$ split into an isolated state
\begin{equation}	      
|T_z\rangle = \tfrac{i}{\sqrt2}(|\!+\!1,-1\rangle-|\!-\!1,+1\rangle) 
\end{equation}
independent on $\Delta/\zeta$ and the degenerate pair
\begin{equation}
|T_{\pm 1}\rangle= \pm |0,\pm 1\rangle\sin\vartheta_1\mp |\pm 1,0\rangle\cos\vartheta_1
\end{equation}
that we rearrange into Cartesian combinations
\begin{align}
|T_x\rangle &=\tfrac{i}{\sqrt{2}}(|T_{+1}\rangle-|T_{-1}\rangle) \,, \\
|T_y\rangle &=\tfrac{1}{\sqrt{2}}(|T_{+1}\rangle+|T_{-1}\rangle) \,.
\end{align}
Here the auxiliary angle $\vartheta_1$ fulfills the equation
$\tan\vartheta_1 = 1/(\sqrt{1+\delta^2}-\delta)$.
A practical advantage of the Cartesian $T_{x,y,z}$ is their direct link to the
components of the magnetic moment to be discussed later.
The remaining five states corresponding to $J=2$ quintuplet at $\Delta/\zeta=0$ 
are at high energies, completely omitted from our model and we thus do not
list them explicitly here.

The energies of the relevant ionic states obtained by diagonalizing
$\mathcal{H}_\mathrm{SOC}+\mathcal{H}_\mathrm{CF}$
\begin{align}
E_s/\zeta &= -\tfrac14\left[1+\sqrt{(1-2\delta)^2+8}\right] -\tfrac16\delta \,, \\
E_{T_z}/\zeta &= -\tfrac12 + \tfrac13\delta \,, \\
E_{T_{x,y}}/\zeta &= -\tfrac12\sqrt{1+\delta^2} -\tfrac16\delta
\end{align}
are plotted in Fig.~\ref{fig:SMlevels}(a) relative to the ground-state energy
$E_s$. As Fig.~\ref{fig:SMlevels}(a) suggests, the selection of $T$ states to be
included in the local basis for a low-energy magnetic model depends on the
$\Delta/\zeta$ ratio. 
In our regime of interest, $\Delta\lesssim-\zeta$, the state $T_z$ is singled
out as the first excited state and together with $s$ it forms a~reduced
two-dimensional basis that can be captured by a pseudospin-$\frac12$.
For roughly $|\Delta|\lesssim\zeta$, one needs to treat all $T$ states on an
equal footing, leading to the four-dimensional basis spanned by $s$, $T_x$,
$T_y$, $T_z$.  We have actually used this largest basis in most calculations
in the paper for the purpose of checking the validity of the simplified
pseudospin-$\frac12$ model in the regime $\Delta\lesssim-\zeta$.
Finally, for $\Delta\gtrsim+\zeta$, $T_z$ is quickly lifted up and the three
remaining states $s$, $T_x$, $T_y$ are sufficient to capture the low-energy
magnetism. In this case, they can be mapped to a~pseudospin-$1$.

Before discussing the two pseudospin regimes in detail, let us specify the
operators of the magnetic moment carried by the four states $s$, $T_\alpha$
($\alpha=x,y,z$). These operators were heavily used in our calculations to
characterize the magnetic phases in terms of static magnetic structure and
magnetic excitation spectra. They are obtained by evaluating the matrix
elements of the general magnetic moment operator $\vc{M}=2\vc S-\vc L$ between
the four ionic states (note the negative sign of effective $t_{2g}$ orbital
momentum contribution as compared to the real $\vc L$). Expressed via the
corresponding creation and annihilation operators, the magnetic moment
components take the form
\begin{align}
M_x &= \gamma_{xy}\, (-i)(s^\dagger T^{\phantom{\dagger}}_x - T^\dagger_x s)
      +\gamma'_{xy}\, (-i)(T^\dagger_y T^{\phantom{\dagger}}_z - T^\dagger_z T^{\phantom{\dagger}}_y) \,, \label{eq:SMmx}\\
M_y &= \gamma_{xy}\, (-i)(s^\dagger T^{\phantom{\dagger}}_y - T^\dagger_y s)
      +\gamma'_{xy}\, (-i)(T^\dagger_z T^{\phantom{\dagger}}_x - T^\dagger_x T^{\phantom{\dagger}}_z) \,, \label{eq:SMmy}\\
M_z &= \gamma_z\, (-i)(s^\dagger T^{\phantom{\dagger}}_z - T^\dagger_z s)
      +\gamma'_z\, (-i)(T^\dagger_x T^{\phantom{\dagger}}_y - T^\dagger_y T^{\phantom{\dagger}}_x) \,. \label{eq:SMmz}
\end{align}
Each component $M_\alpha$ has a Van Vleck-type contribution residing on a
transition between $s$ and $T_\alpha$ states, and a~contribution originating
solely within the $T$-sector. The prefactors entering $M_\alpha$ read as
\begin{align}
\gamma_{xy} &= \cos\vartheta_0 (\sqrt 2 \cos\vartheta_1 + \tfrac1{\sqrt 2} \sin\vartheta_1) 
 + \sin\vartheta_0 (\cos\vartheta_1 + 2\sin\vartheta_1) \,, \\
\gamma'_{xy} &= \sqrt 2 \cos\vartheta_1 - \tfrac1{\sqrt 2} \sin\vartheta_1 \,, \\
\gamma_z &= 3\cos\vartheta_0 \,, \\
\gamma'_z &= 2-3\cos^2\vartheta_1
\end{align}
and are plotted in Fig.~\ref{fig:SMlevels}(b) for the full range of
$\Delta/\zeta$. As expected, in the cubic limit $\Delta/\zeta=0$ where there
is no preference among the $T$ states, the factors partially coincide
\begin{equation}
\Delta/\zeta=0:\qquad
\gamma_{xy}=\gamma_z = \sqrt 6 
\quad\text{and}\quad
\gamma'_{xy}=\gamma'_z=\frac12
\end{equation}
so that the magnetic moment components get unified to
\begin{equation}
\vc M = \sqrt{6}\, (-i)(s^\dagger \vc T - {\vc T}^\dagger s) 
       +\tfrac12\, (-i)(\vc T^\dagger \times \vc T) \,.
\end{equation}

\subsection*{Negative $\Delta$: pseudospin-$\frac12$ case}\label{sec:SMpseudo12}

To capture the low-energy magnetic behavior in the $\Delta\lesssim-\zeta$
regime, it is sufficient to use the two-dimensional basis consisting of $s$
and $T_z$ states. One of the fundamental parameters of the resulting model is
the energy separation of these states, i.e. energy of $T_z$ relative to $s$
which evaluates to
\begin{equation}\label{eq:SMETps12}
E_T = E(T_z)-E(s) = 
\tfrac14\zeta\left[\sqrt{(1-2\delta)^2+8}-(1-2\delta)\right]
= \frac{\zeta}{1-2\delta} \, \frac{2}{1+\sqrt{1+\frac{8}{(1-2\delta)^2}} }
\approx \frac{\zeta}{1-2\delta} \,.
\end{equation}
States from the two-dimensional space spanned by $s$ and $T_z$ can be
described by pseudospin $\widetilde{S}=\frac12$ which can be introduced in a
number of ways. A particularly convenient one that makes the magnetic model
most transparent is based on pseudospin-up and down states defined as the
linear combinations
\begin{align}
|\widetilde{\uparrow}\rangle   &= \tfrac1{\sqrt 2}(|s\rangle+i|T_z\rangle) 
= c^2 |\!-\!1,+1\rangle -s^2 |\!+\!1,-1\rangle - \sqrt2\, cs\, |0,0\rangle \,, 
\label{eq:SMpsup}
\\
|\widetilde{\downarrow}\rangle &= \tfrac1{\sqrt 2}(|s\rangle-i|T_z\rangle) 
= c^2 |\!+\!1,-1\rangle -s^2 |\!-\!1,+1\rangle - \sqrt2\, cs\, |0,0\rangle
\label{eq:SMpsdn}
\end{align}
where $c=\cos\tfrac{\vartheta_0}2$ and $s=\sin\tfrac{\vartheta_0}2$.
The pseudospin operators are then connected to the $s$ and $T_z$ operators as
follows:
\begin{equation}\label{eq:SMpsop}
\widetilde{S}_x = \tfrac12 -n_{T_z} \,, \qquad
\widetilde{S}_y = \tfrac12(s^\dagger T_z + T_z^\dagger s) \,, \qquad
\widetilde{S}_z = -\tfrac{i}2(s^\dagger T_z - T_z^\dagger s) \,.
\end{equation}
In this formulation, the energy cost of having $T_z$ on a given site which is
measured by $E_T n_{T_z}$ with $n_{T_z}=T^\dagger_z T_z$ translates to a 
transverse-field term $-E_T \widetilde{S}_x$. The magnetic moment carried 
by the pseudospin $\widetilde{S}$ is easily obtained by dropping all the 
$T_{x,y}$ terms in \eqref{eq:SMmx}--\eqref{eq:SMmz} which leaves us with
\begin{equation}
M_x =0 \,, \qquad
M_y =0 \,, \qquad
M_z =\gamma_z\, (-i)(s^\dagger T^{\phantom{\dagger}}_z - T^\dagger_z s) \,.
\end{equation}
A comparison with $(M_x,M_y,M_z)=(g_{xy} \widetilde{S}_x, g_{xy}
\widetilde{S}_y, g_z \widetilde{S}_z)$ yields zero in-plane $g$-factor
$g_{xy}=0$ and large out-of-plane $g_z=2\gamma_z$ reaching the value of $6$ in
$\Delta\rightarrow -\infty$ limit.

\subsection*{Positive $\Delta$: pseudospin-$1$ case}

Though it is not directly used in the main text, for completeness we also
describe the mapping of the basis $s$, $T_x$, $T_y$ relevant at large positive
$\Delta/\zeta$ to pseudospin-$1$. In this case, the following relations link
pseudospin-$1$ operators and $s$, $T_{x,y}$ operators:
\begin{equation}\label{eq:SMps1op}
\widetilde{S}_x = -i(s^\dagger T_x - T_x^\dagger s) \,, \qquad
\widetilde{S}_y = -i(s^\dagger T_y - T_y^\dagger s) \,, \qquad
\widetilde{S}_z = -i(T_x^\dagger T^{\phantom{\dagger}}_y - T_y^\dagger T^{\phantom{\dagger}}_x) \,.
\end{equation}
The energy splitting between $s$ and $T_{x,y}$ states expressed via 
$E_T (n_{T_x}+n_{T_y})$ with
\begin{equation}\label{eq:SMETps1}
E_T= E(T_{x,y})-E(s) = \tfrac14\zeta\left[1+\sqrt{(1-2\delta)^2+8}-2\sqrt{1+\delta^2}\right]
\end{equation}
translates now to a single-ion anisotropy term $E_T \widetilde{S}_z^2$. By
removing $T_z$ parts in \eqref{eq:SMmx}--\eqref{eq:SMmz}, we get the projected
magnetic moment operators
\begin{equation}
M_x = \gamma_{xy}\, (-i)(s^\dagger T^{\phantom{\dagger}}_x - T^\dagger_x s) \,, \qquad
M_y = \gamma_{xy}\, (-i)(s^\dagger T^{\phantom{\dagger}}_y - T^\dagger_y s) \,, \qquad
M_z = \gamma'_z\, (-i)(T^\dagger_x T^{\phantom{\dagger}}_y - T^\dagger_y T^{\phantom{\dagger}}_x) \,.
\end{equation}
that immediately give the pseudospin-1 $g$-factors $g_{xy}=\gamma_{xy}$
and $g_z=\gamma'_z$. Both of them approach the value of $2$ in the
$\Delta\rightarrow +\infty$ limit where the orbital contribution to the
magnetic moment is fully quenched and the pseudospin corresponds to a~pure
spin-$1$.


\section{Parameters of the pseudospin-$\frac12$ model}\label{sec:SMexchexpr}

The pseudospin-$\frac12$ model has been derived by evaluating the exchange
interactions originating from nearest-neighbor electronic hopping utilizing
standard second-order perturbation theory. The initial and final bond states
were composed from pseudospin-$\frac12$ states \eqref{eq:SMpsup} and
\eqref{eq:SMpsdn} discussed in Sec.~\ref{sec:SMpseudo12}. The energies of the
virtual states were evaluated using the Hubbard-Kanamori Hamiltonian 
including the usual intra and interorbital Coulomb interaction
($U$, $U'=U-2J_H$) and Hund's coupling ($J_H$) terms
\begin{equation}\label{eq:SMHCoul}
\mathcal{H}_\mathrm{Coulomb}=
U \sum_\alpha n_{\alpha\uparrow} n_{\alpha\downarrow}
\;+\;\sum_{\alpha<\beta} (U'-\tfrac12 J_\mathrm{H})\, n_\alpha n_\beta
-2\sum_{\alpha<\beta} J_\mathrm{H}\, \vc S_\alpha\cdot \vc S_\beta
\;+\;\sum_{\alpha\neq\beta} J_\mathrm{H}\,
\alpha^\dagger_\uparrow\alpha^\dagger_\downarrow
\beta^{\phantom{\dagger}}_\downarrow\beta^{\phantom{\dagger}}_\uparrow
\end{equation}
and neglecting the effects of
$\zeta$ and $\Delta$. Only purely $t_{2g}$ electron configurations were
included in the virtual states. In Eq.~\eqref{eq:SMHCoul}, $\alpha$, $\beta$
run through the $t_{2g}$ orbitals and $n_{\alpha\sigma}$
($\sigma=\uparrow,\downarrow$), $n_{\alpha}=\sum_\sigma n_{\alpha\sigma}$
count the corresponding electron occupations.

In the case of $180^\circ$ bond geometry, the hopping involves two active
orbitals selected by the bond direction and connected diagonally. For the
bonds along $x$ and $y$ direction present in the square lattice, we have (spin
summation is implied):
\begin{equation}
\mathcal{H}^{(x)}_t = -t\, (d^\dagger_{xy}d^{\phantom{\dagger}}_{xy}
+d^\dagger_{zx}d^{\phantom{\dagger}}_{zx})_{ij} + \mathrm{H.c.} 
\,,\qquad
\mathcal{H}^{(y)}_t = -t\, (d^\dagger_{xy}d^{\phantom{\dagger}}_{xy}
+d^\dagger_{yz}d^{\phantom{\dagger}}_{yz})_{ij} + \mathrm{H.c.}
\end{equation}
In the case of $90^\circ$ bond geometry encountered for honeycomb and
triangular lattice, two complementary nearest-neighbor hopping channels need
to be considered. For convenience, we use here the octahedral reference frame
instead of the one shown in Fig.~1(e). On a $z$-bond which is defined by a
metal$_2$-O$_2$ plaquette perpendicular to the $z$~axis (see e.g. Fig.~5 of
Ref.~\cite{Tak21} for a sketch of the hopping geometry), the hopping reads as
\begin{equation}
\mathcal{H}^{(z)}_{tt'} = t\, (d^\dagger_{zx}d^{\phantom{\dagger}}_{yz}
+d^\dagger_{yz}d^{\phantom{\dagger}}_{xz})_{ij}
-t'\,(d^\dagger_{xy}d^{\phantom{\dagger}}_{xy})_{ij}
+ \mathrm{H.c.} 
\end{equation}
The hopping Hamiltonians for the other two bond directions are obtained by
cyclic permutation among $x$, $y$, $z$.

The derivation for both bond geometries gives the following form of the
pseudospin-$\frac12$ model with bond-independent interactions [we are now
switching back to the reference frame of Fig.~1(e)]
\begin{equation}\label{eq:SMpsHam}
\mathcal{H}_{\widetilde{S}}=
\sum_{\langle ij\rangle} \left(
J_z \widetilde{S}^z_i \widetilde{S}^z_j 
+ J_x \widetilde{S}^x_i \widetilde{S}^x_j 
+ J_y \widetilde{S}^y_i \widetilde{S}^y_j \right)
-(h+\delta h) \sum_i \widetilde{S}^x_i \,.
\end{equation}
The expressions for the exchange constants $J_\alpha$ and the correction
$\delta h$ to the transverse field $h=E_T$ can be cast into a universal
structure emphasizing the CF splitting dependence via the auxiliary angle
$\vartheta_0$ entering \eqref{eq:SMpsup} and \eqref{eq:SMpsdn}:
\begin{align}
J_z &= \frac1{U_\mathrm{red}} \,(A_1+A_2\sin^2\vartheta_0) \,, \label{eq:SMJz} \\
J_y &= \frac1{U_\mathrm{red}} \,A_3\sin^2\vartheta_0 \,, \label{eq:SMJy} \\
J_x-J_y &= \frac1{U_\mathrm{red}} \,A_4\sin^4\vartheta_0 \,, \label{eq:SMdJxJy} \\
\delta h &= \frac1{U_\mathrm{red}} \,(A_5+A_6\sin^2\vartheta_0)\sin^2\vartheta_0 \,, \label{eq:SMdh}
\end{align}
where we have adopted a shorthand notation 
$U_\mathrm{red}=U(1-3\eta)(1+2\eta)$ and $\eta=J_\mathrm{H}/U$.
In the $\Delta\lesssim -\zeta$ regime, the factor $\sin^2\vartheta_0$ is
small, vanishing in the $\Delta\rightarrow -\infty$ limit and its asymptotic
$\Delta$-dependence is captured by an expansion
$\sin^2\vartheta_0=2\varepsilon^2-12\varepsilon^4+\mathcal{O}(\varepsilon^6)$
with the small parameter $\varepsilon=1/(1-2\delta)$. The quantities $A_1$ to
$A_6$ containing various combinations of the hopping amplitudes and $\eta$
measuring the relative strength of Hund's coupling are listed below. Note
that being collected at the bonds attached to a given site, the exchange
correction $\delta h$ to the local term depends on the number of nearest
neighbors $z$ for the given lattice.
\vskip 2mm
\noindent\textbullet~$180^\circ$ bonds (square lattice with $z\!=\!4$)
\begin{align}
A_1 &= (5-12\eta-19\eta^2) \, t^2 \,, \\
A_2 &= -4(1-3\eta-5\eta^2) \, t^2 \,, \\
A_3 &= (1+2\eta-5\eta^2) \, t^2 \,, \\
A_4 &= \tfrac14 (1+12\eta+15\eta^2) \, t^2 \,, \\
A_5 &= -(1-6\eta-11\eta^2) \, t^2 \,, \\
A_6 &= -\tfrac12 (1+12\eta+15\eta^2) \, t^2 \,.
\end{align}
\textbullet~$90^\circ$ bonds ($A_{5,6}$ apply to the honeycomb lattice with
$z\!=\!3$,
they are two times larger for the $z\!=\!6$ triangular lattice)
\begin{align}
A_1 &= \tfrac49 [ (7-27\eta-47\eta^2) \, t^2 
     + 2(1+3\eta+7\eta^2) \, t t' + 4(1-3\eta-5\eta^2) \, t'^2 ] \,, \\
A_2 &= -\tfrac23 [ (5-18\eta-31\eta^2) \, t^2
     + (1+4\eta+9\eta^2) \, t t' + 2(1-4\eta-7\eta^2) \, t'^2 ] \,, \\
A_3 &= -\tfrac23 (1+2\eta-5\eta^2) \, t(t+t') \,, \\
A_4 &= (1+3\eta-3\eta^2) \, t^2 - 2\eta(1+2\eta) \, t t' \,, \\
A_5 &= (1+3\eta-3\eta^2) \, t^2 + (1-9\eta^2) \, t t' \,, \\
A_6 &= -\tfrac32 (1+3\eta-3\eta^2) \, t^2 + 3\eta(1+2\eta) \, t t' \,.
\end{align}


\section{Underlying hardcore boson models, mapping to pseudospin-$\frac12$ model}

The full $s$-$\vc T$ model for arbitrary $\Delta$ was obtained by the same
type of calculation as described in Sec.~\ref{sec:SMexchexpr} but considering
now all combinations of four local basis states $|s\rangle$,
$|T_\alpha\rangle$ ($\alpha=x,y,z$) on a bond. This amounts to connecting 16
initial bond states to 16 final bond states by the exchange Hamiltonian to be
constructed, giving $256$ possible bond processes in total. The model
Hamiltonian is too complex to be presented explicitly here, but its general
structure is of the form
\begin{equation}\label{eq:SMsTHam}
\mathcal{H}_{s\text{-}\vc T} = 
\sum_i E_m 
\Psi^\dagger_{m i}
\Psi^{\phantom{\dagger}}_{m i}
+
\sum_{\langle ij\rangle} 
V_{mm'nn'} 
\Psi^\dagger_{m i}
\Psi^\dagger_{m' j}
\Psi^{\phantom{\dagger}}_{n i}
\Psi^{\phantom{\dagger}}_{n' j}
\end{equation}
involving four-component vector bosons $\vc \Psi=(s,T_x,T_y,T_z)$ subject to
local hardcore constraint $n_s+n_{T_x}+n_{T_y}+n_{T_z}=1$ at each site (here
$n_s=s^\dagger s$, $n_{T_x}=T_x^\dagger T_x^{\phantom{\dagger}}$ etc.).
Summation over indices $m,m',n,n'=0,1,2,3$ is implied. The on-site excitation
energies $E_m$ are given by the multiplet levels as in \eqref{eq:SMETps12} and
\eqref{eq:SMETps1}, potentially supplemented by negligible exchange
corrections. The amplitudes $V_{mm'nn'}$ are functions of $t$, $t'$, $U$,
$J_H$ and the auxiliary angles $\vartheta_0$, $\vartheta_1$. The corresponding
expressions would be similar to those presented in Sec.~\ref{sec:SMexchexpr}
but more complex. Moreover, they depend on the direction of the particular
bond $\langle ij\rangle$ since the bond-directional states $T_x$, $T_y$ are
included.

In the $\Delta\lesssim-\zeta$ regime, it is sufficient to employ a much
simpler hardcore boson model obtained by a projection of \eqref{eq:SMsTHam}
onto $s$, $T_z$ subspace. The exchange interactions in this reduced model are
fully captured by only a few terms -- hopping of $T_z$ particles, their
pairwise creation and annihilation, and their repulsion or attraction.
Including a small correction to $s$-$T_z$ splitting $E_T$ due to exchange
processes, the resulting Hamiltonian reads as
\begin{equation}\label{eq:SMsTzHam}
\mathcal{H}_{s\text{-}T_z} = 
\sum_{\langle ij\rangle} \Bigl\{
\tau\, \bigl[ (T_z^\dagger s)_i (s^\dagger T_z^{\phantom{\dagger}})_j + \text{H.c.} \bigr] 
-\kappa\, \bigl[ (T_z^\dagger s)_i (T_z^\dagger s)_j + \text{H.c.} \bigr]
+ V\, n_{T_z i}\, n_{T_z j} \Bigr\}
+(E_T+\delta E_T)\sum_i n_{T_zi} \,.
\end{equation}

The parameters $\tau$, $\kappa$, $V$ are independent on the bond direction
since we projected out the states $T_{x,y}$ having bond-directional wave
functions. By rewriting the terms of the above Hamiltonian using a new basis
composed of 
$|\widetilde{\uparrow}\rangle = \tfrac1{\sqrt 2}(|s\rangle+i|T_z\rangle)$
and 
$|\widetilde{\downarrow}\rangle = \tfrac1{\sqrt 2}(|s\rangle-i|T_z\rangle)$,
the $s$-$T_z$ model can be mapped to the pseudospin-$\frac12$ model
\eqref{eq:SMpsHam}. This way one finds the relations connecting the two
formulations in terms of parameters
\begin{equation}\label{eq:JparsfromsTZ}
J_x = V
\,,\quad
J_y = 2(\tau - \kappa)
\,,\quad
J_z = 2(\tau + \kappa)
\,,\quad
\delta h = \delta E_T +\tfrac12 zV 
\end{equation}
and giving identical expressions for $J_{x,y,z}$ and $\delta h$ as 
\eqref{eq:SMJz}--\eqref{eq:SMdh}
obtained by working out the second order perturbation theory directly in the
$|\widetilde{\uparrow}\rangle$, $|\widetilde{\downarrow}\rangle$ basis.


\section{Phase diagrams and excitation spectra at a harmonic level}

To establish approximate phase diagrams of the above hardcore boson models,
we use a simple variational approach based on factorized trial states.
For the full $s$-$\vc T$ model \eqref{eq:SMsTHam} or the projected 
$s$-$T_z$ model \eqref{eq:SMsTzHam} [equivalent to the pseudospin-$\frac12$
model \eqref{eq:SMpsHam}], we use
\begin{equation}\label{eq:SMPsitrial}
|\mathrm{trial}\rangle = \prod_i \Bigl(\sqrt{1-\rho_i}\; s^\dagger 
+ \sqrt{\rho_i}\; {\textstyle \sum_\alpha} 
d^*_{i\alpha} T_{i\alpha}^\dagger \Bigr) \; |\mathrm{vac}\rangle
\qquad\text{or}\qquad
\prod_i \Bigl(\sqrt{1-\rho_i}\; s^\dagger 
+ \sqrt{\rho_i}\;\mathrm{e}^{i\phi_i} T_{iz}^\dagger \Bigr) \; |\mathrm{vac}\rangle 
\,,
\end{equation}
respectively. These enable to assess the tendency of triplets to form a
condensate and to determine the preferred condensate structure. We consider
unit cells of various sizes and minimize the average value of the model
Hamiltonian with respect to position-dependent variational parameters $\rho_i$
(``condensate density'') and either the complex vectors $\vc d_i$ embedding
direction and phases in the triplet space ($s$-$\vc T$ model) or just the
phases $\phi_i$ of $T_z$ ($s$-$T_z$ model). The resulting optimal
configuration can be used to evaluate e.g. the magnetic structure via
\eqref{eq:SMmx}--\eqref{eq:SMmz} or pseudospin structure via
\eqref{eq:SMpsop}.

To determine the excitation spectra on a harmonic expansion level, we have
adopted a variant of linear flavor wave theory (LFWT)
\cite{Pap84,Pap88,Chu90,Jos99,Som01} that proceeds according to the following
recipe relevant for the $s$-$\vc T$ model case. First, using the result of
the previous minimization, a sublattice-dependent rotation among the bosonic
operators $(s,T_x,T_y,T_z)\rightarrow(c,a,b_1,b_2)$ is introduced to bring the
optimal $|\mathrm{trial}\rangle$ state to a simplified form
\begin{equation}
|\mathrm{trial}\rangle = \prod_i c^\dagger_i \;|\mathrm{vac}\rangle \,.
\end{equation}
This requirement implies the first part of the transformation -- the new boson
$c$ needs to be defined via the relation
$c=\sqrt{1-\rho}\; s + \sqrt{\rho}\; {\textstyle \sum_\alpha} d_{\alpha} T_{\alpha}$.
Boson $a$ is an orthogonal complement of $c$ given by
$a=\sqrt{\rho}\; s - \sqrt{1-\rho}\; {\textstyle \sum_\alpha} d_{\alpha} T_{\alpha}$.
It is constructed to maintain the $\vc d$ structure and as such it is linked
to an amplitude oscillation of the condensate. The remaining two bosons $b_1$
and $b_2$ are constructed by continuing the orthogonalization procedure which
generates directions perpendicular to $\vc d$ in the triplet space:
$b_n=\sum_\alpha d^{(n)}_\alpha T_\alpha$ ($n=1,2$) with $\vc d^{(n)}\cdot 
\vc d=0$ and $\vc d^{(1)}\cdot \vc d^{(2)}=0$. Therefore, $b_1$ and $b_2$ carry
two magnon-like modes altering primarily the magnetic structure of the
condensate and not its amplitude. After the rotation is performed, the
condensed combination of boson operators is replaced as
$c,c^\dagger\rightarrow \sqrt{1-n_a-n_{b_1}-n_{b_2}}$ to approximately account
for the hardcore constraint $n_s+n_T = n_c+n_a+n_b = 1$ and the resulting
Hamiltonian is expanded up to second order in boson operators providing a
harmonic Hamiltonian involving three bosons $a$, $b_1$, $b_2$ per site in the
unit cell. As the last step, this Hamiltonian is diagonalized via multiboson
Bogoliubov transformation. The procedure for the $s$-$T_z$ model is simpler
as it does not involve the two bosons $b_1$ and $b_2$. In principle, thanks to
the equivalence of the two formulations \eqref{eq:SMsTzHam} and
\eqref{eq:SMpsHam}, the excitation spectrum obtained within $s$-$T_z$ model
could be also accessed by solving the equivalent pseudospin-$\frac12$ model.
However, if the pseudospin-$\frac12$ model is treated by linear spin-wave
theory, there are slight quantitative differences since the constraints are
handled differently and the two harmonic expansions are not identical. 

Finally, the harmonic modes obtained using the above procedure and the
associated matrix propagator can be used to evaluate the dynamic magnetic
susceptibility
\begin{equation}
\chi_{\alpha\alpha}(\vc q,\omega) = i \int_0^\infty 
\langle\, [ M_\alpha(\vc q,t) M_\alpha(-\vc q,0) ] \,\rangle 
\,\mathrm{e}^{i(\omega+i0^+)t} \,\mathrm{d}t 
\end{equation}
presented in the main text and the following section. Via the above bosonic
rotation and a subsequent substitution for $c$, $c^\dagger$, the magnetic
moment operator $M_\alpha$ can be expressed via the $a$, $b_1$, and $b_2$
operators. Keeping only the linear terms in $M_\alpha$, the susceptibility can
be constructed from the components of the matrix propagator for the $a$,
$b_1$, and $b_2$ bosons.


\section{Additional magnetic susceptibility spectra}

\subsection*{Square lattice}

\begin{figure*}[htb]
\includegraphics[scale=1.06]{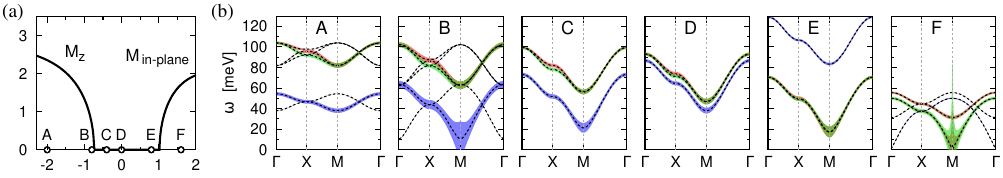}
\caption{Sample magnetic excitation spectra covering both the negative and
positive $\Delta$ parts of the phase diagram in Fig.~2(a) of the
main text [reproduced in (a)]. One can clearly see the distinct fate of the
$xx$, $yy$ (red and green) and $zz$ (blue) modes on the two sides of the
$\Delta=0$ point D, where they are quasidegenerate. While flat Ising-type
excitations are observed on the left for $\Delta<0$ after the AF order
develops (point A), at $\Delta>0$ on the right, $T_z$ is quickly lifted up and
the low-energy behavior is captured by effective spin-1 model where exchange
interactions of predominantly XY-type compete with varying single-ion
anisotropy \cite{Jai17,Tak21}. In the ordered regime it features nearly
gapless magnons and a gapful amplitude mode (point F). The approaching QCP is
generally signaled by a characteristic soft mode that touches zero energy at
QCP (near points B and~E).
}\label{fig:SMexc_sqr}
\end{figure*}

Figure~\ref{fig:SMexc_sqr} presents sample magnetic excitation spectra
obtained for the same parameter setup as in Fig.~2 of the main
text. In addition to the three spectra for the A, B, C points with $\Delta<0$
that were already shown in Fig.~2, here we cover also the positive
$\Delta$ range to contrast the nature of magnetic excitations in these two
cases. For $\Delta\gtrsim +\zeta$, the low-energy magnetic behavior can be
roughly described by pseudospin-$1$ model
\begin{equation}\label{eq:SMps1Ham}
\mathcal{H}_{\widetilde{S}}=
\sum_{\langle ij\rangle} \left[
J_{xy} \left( \widetilde{S}^x_i \widetilde{S}^x_j +
\widetilde{S}^y_i \widetilde{S}^y_j \right)
+J_z \widetilde{S}^z_i \widetilde{S}^z_j \right]
+E_T \sum_i (\widetilde{S}^z_i)^2 
\end{equation}
where the pseudospin-$1$ operators have been introduced in \eqref{eq:SMps1op}.
The large single-ion anisotropy term pushes the pseudospins into the $xy$
plane giving them specific dynamics manifesting itself by XY-type magnon
dispersion \cite{Jai17}.

The spectra shown in Fig.~\ref{fig:SMexc_sqr}(b) are calculated using the LFWT
for the $s$-$\vc T$ model. If the LFWT is applied to the reduced models,
either $s$-$T_z$ (equivalent to pseudospin-$\frac12$ model) or $s$-$T_x$-$T_y$
(equivalent to pseudospin-$1$ model), the corresponding parts of the magnetic
excitation spectra are exactly reproduced. This applies to the phases where
the complementary bosons do not participate in condensation. Hence, on the
level of LFWT, the $s$-$T_z$ model fully captures the $zz$ susceptibility in
the PM phase and the left AF phase, covering the entire range of interest on
both sides of the QCP. Similarly, the $s$-$T_x$-$T_y$ model exactly
reproduces the $xx$ and $yy$ susceptibility in the PM phase and the right AF
phase.

\subsection*{Honeycomb lattice}

\begin{figure*}[b]
\includegraphics[scale=1.06]{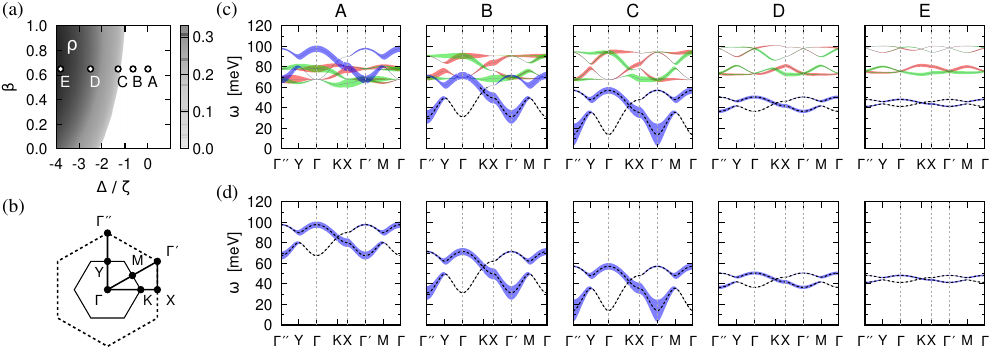}
\caption{Magnetic excitation spectra for a set of points in the honeycomb
phase diagram of Fig.~3(c), plotted in (a) using the condensate
density $\rho$.
(b)~Brillouin zone of the honeycomb lattice (solid) and the triangular lattice
obtained by filling the hexagon ``voids'' (dashed). High-symmetry points and
the path used to plot the spectra are indicated. The outer hexagon coincides
with the triangular Brillouin zone used in Figs.~4,
\ref{fig:SMexc_tri_clock}, \ref{fig:SMexc_tri_stripe}.
(c)~Magnetic susceptibility obtained by applying LFWT to the full
\mbox{$s$-$\vc T$} model. The individual components are shown in red ($xx$),
green ($yy$), and blue ($zz$) with the line thickness indicating the
intensity.
(d)~$zz$~component of the magnetic susceptibility obtained using the projected
$s$-$T_z$ model \eqref{eq:SMsTzHam} which is equivalent to the
pseudospin-$\frac12$ model \eqref{eq:SMpsHam} (TFIM extended by $J_{x,y}$
interactions) with the parameters given by Eq.~\eqref{eq:JparsfromsTZ}. The
remaining $xx$ and $yy$ component are zero due to vanishing in-plane
$g$-factors.
}\label{fig:SMexc_hon}
\end{figure*}

The excitation spectra complementing the $\Delta<0$ phase diagram of the
honeycomb model presented in Fig.~3(c) of the main text are shown
in Fig.~\ref{fig:SMexc_hon}. Since the honeycomb lattice does not suffer from
the geometric frustration, the picture is rather simple. As we drive the
system from PM to AF phase, the excitations first soften at the AF momentum
$\vc q=\Gamma'$, harden back after passing through the QCP, and finally become
Ising-like with a~flat dispersion as we get deeper into the AF phase. The
overall behavior is thus very similar to that observed in the previous
paragraph for the square lattice.
A comparison of Fig.~\ref{fig:SMexc_hon}(c) and Fig.~\ref{fig:SMexc_hon}(d)
shows that the relevant out-of-plane magnetic modes entering $zz$ component of
the magnetic susceptibility are well captured by the reduced $s$-$T_z$
Hamiltonian equivalent to the pseudospin-$\frac12$ model.

\begin{figure*}[tb]
\includegraphics[scale=1.02]{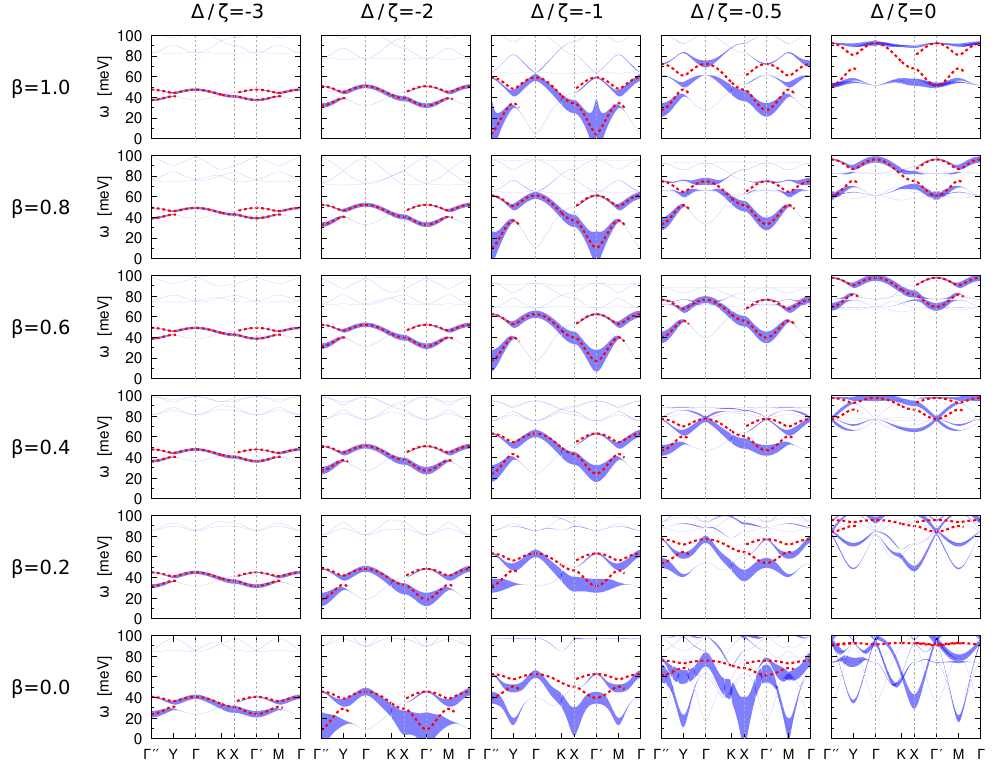}
\caption{Comparison of the $zz$ component of the magnetic susceptibility
obtained by LFWT applied either to the full $s$-$\vc T$ model (blue) or to the
projected $s$-$T_z$ model (red dashed). In the former case, the line thickness
indicates the intensity as before. In the latter case, to give a hint
concerning the intensity, we plot only parts of the dispersion of the
corresponding modes where the intensity exceeds certain threshold. The
combinations of $\Delta/\zeta$ and $\beta$ values cover most of the phase
diagram in Fig.~\ref{fig:SMexc_hon}(a) to give an overall picture.
}\label{fig:SMexc_hon2}
\end{figure*}

This point is elaborated further in Fig.~\ref{fig:SMexc_hon2} where we
systematically cover the phase diagram of Fig.~\ref{fig:SMexc_hon}(a) focusing
solely on the $zz$ component of the magnetic susceptibility with the aim to
test the pseudospin-$\frac12$ model. We can observe a good match between the
results of the $s$-$\vc T$ model and $s$-$T_z$ model for $\Delta\lesssim
-\zeta$, which is the domain of applicability of the reduced model, and the
agreement is rather good even at $\Delta/\zeta=-0.5$ if $t'$ is comparable or
larger than $t$ (i.e. $\beta\geq 0.4$ in our plots). As in the case of the
square lattice, this suggests the availability of TFIM-like physics on both
sides of QCP.
However, in contrast to the square-lattice case, the harmonic $\chi_{zz}$
spectra obtained for $s$-$\vc T$ and $s$-$T_z$ models, respectively, do not
\textit{exactly} coincide in the entire PM phase. As it turns out by an
inspection of the interactions within $s$-$\vc T$ model, there is a bilinear
coupling between $T_z$ and $T_{x,y}$ bosons for $90^\circ$ bond geometry which
causes the difference observed on the level of LFWT. This coupling becomes
negligible around $\beta\approx0.6$ for our parameter setup, leading to a
separation of the modes and the very good agreement observed even at
$\Delta/\zeta=0$, well outside the region governed by pseudospin-$\frac12$
model. Far away from $\beta\approx 0.6$, the difference quite significant in
our plots for $\Delta/\zeta \geq -0.5$. In the case of the square lattice,
such bilinear terms are absent, explaining the perfect match.
While for $\beta$ around $0.6$ the pseudospin-$\frac12$ model can be pushed
surprisingly far beyond its formal range of applicability, in the vicinity of
$\beta=0$ and $\beta=1$ this extension is further prevented by two
complementary kinds of frustration related to a competition among the three
$T$ flavors, all being equally active for $\Delta=0$.
%
%
Near $\beta=1$ and $\Delta=0$, the corresponding singlet-triplet model
features Kitaev-like frustration leading to a peculiar ground state with
strong but nearest-neighbor only correlations \cite{Cha19}. This is also
reflected in the corresponding excitation spectrum in
Fig.~\ref{fig:SMexc_hon2}. 
For $\beta=0$ and $\Delta=0$ and neglected Hund's coupling, Ref.~\cite{Kha13}
found the motion of the triplets of a~given flavor to be restricted to a set
of parallel zigzag chains on the honeycomb lattice, leading to enhanced
quasi-1D behavior. 
With the Hund's coupling included, the picture changes somewhat, but the
treatment based on the trial state \eqref{eq:SMPsitrial} still shows a
tendency to support non-trivial condensate with a 4-sublattice stripy
structure. Such a phase is found in a~very narrow region in the phase diagram
of Fig.~\ref{fig:SMexc_hon}(a) ($\beta\lesssim 0.03$ at $\Delta/\zeta=0$;
$\Delta/\zeta\gtrsim -0.6$ at $\beta=0$). The inspection of the corresponding
excitation spectra reveals the characteristic momenta $M$, $Y$, $X$ as
relevant to this type of condensation.

\subsection*{Triangular lattice}

\begin{figure*}[hb]
\includegraphics[scale=1.06]{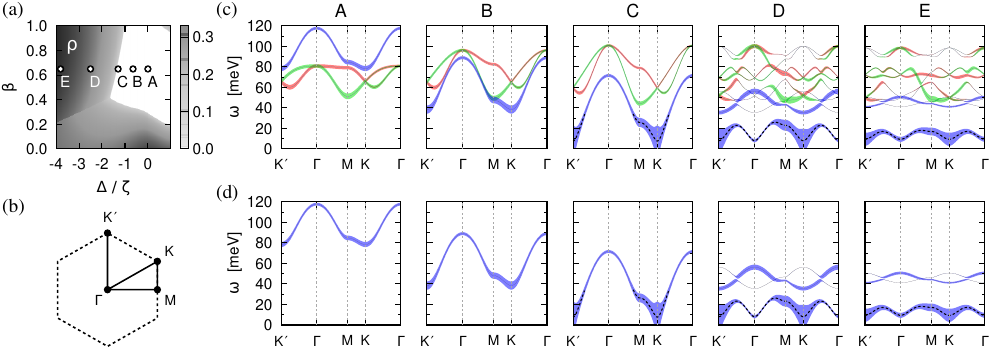}
\caption{The same as in Fig.~\ref{fig:SMexc_hon} for the triangular lattice
and a set of points illustrating the transition from the disordered phase to
the clock phase. Again the reduced pseudospin-$\frac12$ model in the hardcore
boson formulation of Eq.~\eqref{eq:SMsTzHam} perfectly reproduces the major
part of the magnetic excitation spectra.
}\label{fig:SMexc_tri_clock}
\end{figure*}

\begin{figure*}[hb]
\includegraphics[scale=1.06]{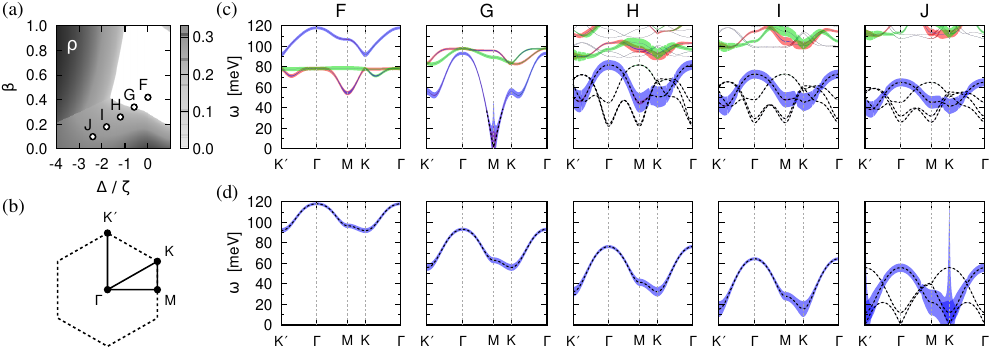}
\caption{The same as in Fig.~\ref{fig:SMexc_hon} for the triangular lattice
and a set of points illustrating the transition from the disordered phase to
the stripe phase. Here the reduced pseudospin-$\frac12$ model is insufficient
since the condensation into stripe phase involves $T_x$ levels in addition to
the $T_z$ ones. Therefore the corresponding phase boundary is not detected in
(d) and only later (near point J) a transition to the clock phase takes place.
However, the full $s$-$\vc T$ model continues to support the stripe phase in
that parameter regime.
}\label{fig:SMexc_tri_stripe}
\end{figure*}

Figures~\ref{fig:SMexc_tri_clock} and \ref{fig:SMexc_tri_stripe} illustrate
the evolution of the magnetic excitations when entering the two condensed
phases encountered in the case of the triangular lattice. The corresponding
phase diagram is shown in Fig.~3(d) of the main text. The spectra
in Fig.~\ref{fig:SMexc_tri_clock} are relevant to the clock phase and were
discussed in the context of Fig.~4 of the main text. As
Fig.~\ref{fig:SMexc_tri_clock} demonstrates, the evolution of the dominant
out-of-plane component of the magnetic susceptibility is again well reproduced
-- in the entire range -- within the reduced pseudospin-$\frac12$ model. 
However, in the case of Fig.~\ref{fig:SMexc_tri_stripe} covering the
transition into the stripe phase, the pseudospin-$\frac12$ model comes out as
insufficient due to the significant interaction between the $T_z$ and
$T_{x,y}$ bosons which makes the latter ones to participate in the
condensation.

\end{document}